\documentclass[preprint]{aastex}
\usepackage{natbib}
\shorttitle{Type II migration of gas giants}

\shortauthors{Ida et al.}

\begin{document}

\title{Slowing Down Type II Migration of Gas Giants to Match Observational Data}

\author{Shigeru Ida$^1$, Hidekazu Tanaka$^2$, Anders Johansen$^{3}$, Kazuhiro D. Kanagawa$^{4,5}$, and Takayuki Tanigawa$^6$}
\affil{1) Earth-Life Science Institute, Tokyo Institute of Technology,
Meguro-ku, Tokyo 152-8550, Japan \\
2) Astronomical Institute, Tohoku University, Aramaki, Aoba-ku, Sendai, Miyagi 980-8578, Japan \\
3) Lund ObservatoryDepartment of Astronomy and Theoretical Physics, Lund University, Box 43, 221 00, Lund, Sweden\\
4) Research Center for the Early Universe, Graduate School of Science, University of Tokyo, Hongo, Bunkyo-ku, Tokyo 113-0033, Japan \\
5) Institute of Physics and CASA$^{\ast}$, 
Faculty of Mathematics and Physics, University of Szezecin, Wielkopolska 15, PL-70-451 Szczecin, Poland \\
6) National Institute of Technology, Ichinoseki College, Takanashi, Hagisho, Ichinoseki, Iwate 021-8511, Japan}
\email{ida@elsi.jp}

\begin{abstract}
The mass and semimajor axis distribution of 
gas giants in exoplanetary systems obtained by radial velocity surveys
shows that super-jupiter-mass planets are piled up at $\ga 1$ au,
while jupiter/sub-jupiter-mass planets are broadly distributed
from $\sim 0.03$ au to beyond 1 au.
This feature has not been explained by theoretical predictions.
In order to reconcile this inconsistency,
we investigate evolution of gas giants with a new type II migration formula 
by \citet{Kanagawa18},
by comparing the migration, growth timescales of gas giants, and disk lifetime,
and by population synthesis simulation.
While the classical migration model assumes that a gas giant opens up a clear gap
in the protoplanetary disk and
the planet migration is tied to the disk gas accretion,
recent high-resolution simulations show that the migration of gap-opening 
planets is decoupled from the disk gas accretion
and \citet{Kanagawa18} proposed that type II migration speed is nothing other than type I migration speed
with the reduced disk gas surface density in the gap.
We show that with this new formula, type II migration is significantly 
reduced for super-jupiter-mass planets,
if the disk accretion is driven by the disk wind as suggested by recent MHD simulations.
Population synthesis simulations show that super-jupiter-mass planets
remain at $\ga 1$ au without any additional ingredient such as disk photoevaporation.
Therefore, the mystery of the pile-up of gas giants at $\ga 1$ au
will be theoretically solved if the new formula is confirmed
and wind-driven disk accretion dominates.   
\end{abstract}

\section{Introduction}

\begin{figure}[htbp]
  \centering
  \includegraphics[width=100mm, angle = 0]{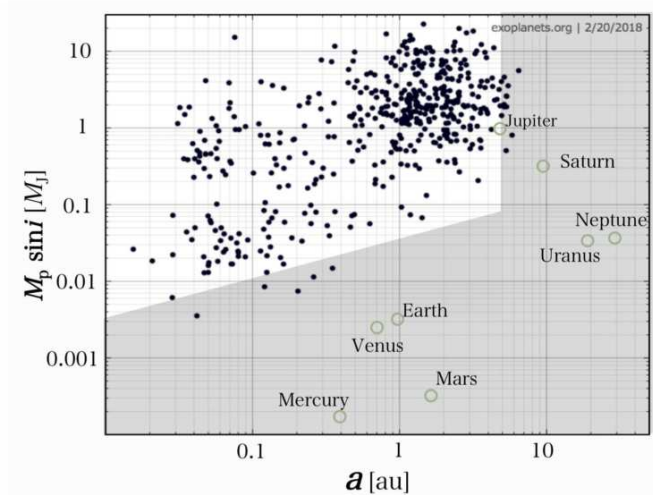}
  \caption{Mass and semimajor axis distribution of exoplanets discovered by the radial velocity survey
  (http://exoplanets.org).  The units of the planetary mass ($M_{\rm p}$) and semimajor axis ($a$) are
  jupiter mass ($M_{\rm J}$) and astronomical unit (au). 
  The shaded region expresses the parameter space
  where it is not easy to detect planets with the current surveys.
  The planets in our solar systems are also plotted as a reference.}
  \label{fig:Ma}
\end{figure}

Radial velocity (RV) surveys show that
more than 10\% of solar-type stars have gas giant planets \citep[e.g.,][]{Cumming08}.
The RV data (Fig.~\ref{fig:Ma}) show that gas giants---in particular, 
super-jupiter-mass planets---are piled up beyond 1 au in exoplanetary systems, 
while 1\% of solar-type stars have ``hot jupiters''
with the semimajor axes $a \la 0.1$ au.
Type II migration is one of the promising mechanisms to account for
the small semimajor axes of hot jupiters \citep[e.g.,][]{Lin96},
although they can also be formed by planet-planet scattering followed by
tidal circularization \citep[e.g.,][]{RasioFord96, Nagasawa08, Winn10}.

In the classical model of type II migration, a giant planet opens up a gap
in the protoplanetary disk and its migration is tied to global disk accretion
\citep[e.g.,][]{Lin1986}. 
It is assumed that disk gas does not cross the gap.
However, in this case, the migration timescale 
is equal to the local viscous diffusion timescale, which
is generally much shorter than the global disk diffusion timescale (disk lifetime).
It is predicted that most of the gas giants
become hot jupiters \citep[e.g.,][]{IL08, Hasegawa13, Bitsch15b}
and it is apparently inconsistent with the RV data.
Figure~\ref{fig:Ma} shows 
the mass and semimajor axis distribution of exoplanets discovered by 
RV surveys. 
To remove strong observational bias toward close-in orbits, the planets discovered 
by transit survey are not included.
This plot clearly shows that most of planets with more than Jupiter mass ($M_{\rm J}$)
are located beyond 1 au and, compared with the population, hot jupiters are rather rare.

In the final phase of disk evolution in which
planets can become more massive than the remaining disk mass, 
the disk cannot sufficiently subtract
the planetary orbital angular momentum and the migration is slowed down  (``planet-dominated'' migration).
However, the contribution of the final phase to the total migration is very limited (Hasegawa \& Ida 2013). 

In the population synthesis simulations by Ida \& Lin 
\citep{IL04, IL04b, IL05, IL08, IL10, IL13}, which will be referred to as ``IL'' in this paper, 
type II migration is assumed to be 10 times slower than the classical model,
taking into account uncertainty, in order to avoid apparent inconsistency
between the predicted semimajor axis distribution and the RV data.
Nevertheless, they still predicted the formation of too many hot jupiters
and the 10 times reduction in type II migration speed is not justified.  

On the other hand, 
in the population synthesis simulations by Alibert, Mordasini, and Benz 
\citep[][referred to as ``AMB"]{AMB09, AMB09b, AMB11, AMB13}
assumed that all the disk accretion flow is accreted by the planet without any impedance by the gap.\footnote{
\citet{Mordasini2012} introduced Bondi and Hill limits \citep{D'Angelo2010}
instead of the disk accretion limit.  However, the additional limits
do not take into account a reduction of the surface density in the accretion band
due to gap opening that we discuss in the present paper.
For simplicity, we call the the unperturbed disk accretion limit 
as the AMB model. }

Then, the migration is linked to the planet growth
and is quickly transferred to the planet-dominated migration,
which avoids the formation of too many hot jupiters,
as shown in section 2.2.
However, in this model, a gas giant in principle accretes most of the initial disk gas
to acquire much larger gas mass than the upper limit of observed planetary masses ($\sim 10 M_{\rm J}$), except when the gas giant forms in a significantly depleted disk.
AMB introduced strong external photoevaporation 
in order to avoid the formation of the very large gas giants.
However, the assumption that the planet accretes 
the full amount of gas flow through the disk would not be justified.

\citet{TT16} (hereafter ``TT") discussed the details of 
the fraction of disk gas flow that is accreted by a planet (the accretion efficiency) and derived
an analytical formula for this.
As discussed in detail in section 2.2, they found that
the efficiency is low for both sufficiently small and large planets.
They also raised the important point that disk gas surface density 
in the disk regions interior to 
the planet orbit is decreased by the gas accretion onto the planet,
in particular Jupiter-mass planets with high accretion efficiency.
According to the depletion of the inner disk,
type II migration turns into the relatively slow planet-dominated regime.
However, for more massive planets, this reduction is not effective,
while the RV data show that more massive planets stay beyond 1 au.

It is most likely that we have missed some important intrinsic physics of type II migration. 
Recently, \citet{Kanagawa18} proposed a new physical interpretation
and formula for type II migration,
combining hydrodynamical simulations with broad ranges of parameters.
In this paper, we discuss the impact of the new formula on the
evolution of gas giants, 
while its validity of the new formula needs to be tested by follow-up simulations.
We demonstrate that the new formula reconciles the inconsistency
between the theoretical predictions and the RV data,
if the global angular momentum transfer of the disk is dominated by
the disk wind rather than the turbulent diffusion.

Recent non-ideal MHD simulations \citep[e.g.,][]{Bai17} show that 
Magneto-Rotational Instability (MRI) 
is suppressed in most parts of the disk due to Ohmic dissipation, the Hall effect, and ambipolar diffusion.
Observation of the HL Tau disk suggests that this is the case \citep{Pinte16, Hasegawa17},
because the oblique image of the gaps in the disk do not show narrow parts due to stirred-up grains.
However, the measured typical gas accretion rate onto classical T-Tauri stars 
is equivalent to $\alpha \sim 10^{-3}-10^{-2}$ in the case of turbulent diffusion (see section 2.1).
Because it is likely that the actual turbulent strength, $\alpha_{\rm vis}$, 
is much smaller than $10^{-3}-10^{-2}$ in the
MRI-inactive disks, it has been proposed that 
the MHD disk wind may be a main driver for global disk mass accretion
\citep{Bai16, Suzuki16}.
The disk angular momentum is carried away by the wind rather than
transferred to outer disk regions through turbulent viscous stress.
The angular momentum transfer induces inward disk gas accretion,
which is called ``wind-driven accretion."
We can define the equivalent alpha parameter, $\alpha_{\rm acc}$,
by the disk wind \citep{Armitage13}.
\citet{Hasegawa17} suggested that $\alpha_{\rm vis}$ may be one order of magnitude smaller
than $\alpha_{\rm acc}$
and the disk accretion rate is determined by $\alpha_{\rm acc}$, rather than $\alpha_{\rm vis}$.

\citet{Kanagawa15} and \citet{Duffell15} showed that the gap depth is deeper
for a smaller $\alpha_{\rm vis}$ and for a larger planetary mass $M_{\rm p}$. 
Because \citet{Kanagawa18} proposes that type II migration is slower for 
a planet with a deeper gap, it is slower for smaller values of $\alpha_{\rm vis}$ and larger $M_{\rm p}$.
On the other hand, disk lifetime is determined by the global disk accretion
and is inversely proportional to $\alpha_{\rm acc}$.
It is expected that large gaseous planets do not undergo significant 
type II migration for $\alpha_{\rm vis} \ll \alpha_{\rm acc}$.\footnote{ 
\citet{Alexander2012}, \citet{Ercolano2015}, and \citet{Jennings2018}
proposed that a clear gap created by internal XUV photoevaporation truncates migrations of gas giants.
Gas giants can be piled up beyond 1 au by this mechanism, if XUV flux is strong enough. 
However, the photoevaporation gap is independent of the planetary mass,
while the model we propose here is the effect of the gap created by
the planetary perturbations and the gap depth is sensitive to
the planetary mass.
The sensitive mass dependence of the pile-up found by the RV observations
would be more consistent with our model, although more detailed
comparison must be done.}

In section 2, we summarize the classical formula of type II migration and its 
problems in reproduction of the observed distributions of gas giants.
In section 3, we show how the new formula slows down the migration of planets
with masses larger than Jupiter mass.
We perform population synthesis simulations with the classical and new formulas
in section 4, 
which show how the new model powerfully solves the inconsistency between
the RV data and the classical model. 
Section 5 is a summary.

\section{Classical model of type II migration}
\subsection{Difficulty in surviving migration}

In the classical model of type II migration,
it is assumed that disk gas does not cross the gap.
Then, the migration timescale of a giant planet located at $r$ is 
simply give by \citep[e.g.,][]{Hasegawa13}
\begin{eqnarray}
t_{\rm mig2,cl} & \sim & \frac{2\pi \Sigma r^2 + M_{\rm p}}{\dot{M}_{\rm g}} 
\sim \frac{2\pi \Sigma r^2 + M_{\rm p}}{3 \pi \Sigma \nu_{\rm acc}} \label{eq:cl} \\
 & = & \frac{2r^2}{3\nu_{\rm acc}} 
 \left(1 +\frac{M_{\rm p}}{2 \pi \Sigma r^2} \right) 
 = \frac{2}{3\Omega} \alpha_{\rm acc}^{-1} \left(\frac{h}{r}\right)^{-2} 
 \left(1 +\frac{M_{\rm p}}{2 \pi \Sigma r^2} \right),  \label{eq:cl2}
 \end{eqnarray} 
where $M_{\rm p}$ is the planet mass, $\Sigma$ is the unperturbed gas surface density 
(outside the gap), $2\pi \Sigma r^2$ is the total disk mass
inside the planetary orbit at radius $r$ (see below), $\dot{M}_{\rm g}$ is 
the disk accretion rate toward the host star, 
$\nu_{\rm acc}$ is the effective kinetic viscosity for disk accretion,
represented by $\nu_{\rm acc} = \alpha_{\rm acc} h^2 \Omega = \alpha_{\rm acc} (h/r)^2 r^2 \Omega$,
$h$ is the disk gas scale height, $\Omega$ is the Kepler frequency,
and $\alpha_{\rm acc}$ is an alpha parameter for disk gas accretion.
If turbulent viscous diffusion is dominant over the disk wind in disk angular momentum transfer,  $\alpha_{\rm acc} = \alpha_{\rm vis}$.
If the disk wind is dominant,
$\alpha_{\rm acc}$ is an "effective" parameter determined by the disk wind
and $\alpha_{\rm acc} > \alpha_{\rm vis}$.

We adopt the self-similar solution for disk evolution \citep{Lynden-Bell7}.
The gas disk surface density in the self-similar solution is given by
\begin{equation}
\Sigma(r,t) = \Sigma_0 \left(\frac{r}{r_0}\right)^{-1} \tilde{t}^{-3/2} 
 \exp \left(-\frac{r}{\tilde{t} r_0}\right),
 \label{eq:SSS}
\end{equation}
where $\tilde{t} = (t/t_{\rm dep}) + 1$ and $r_0$ is the initial disk radius,
the subscript ``0" represents values at $r_0$, and
$t_{\rm dep}$ is the global disk depletion time (disk lifetime).
The initial total disk mass is given by $\int^r 2 \pi r \Sigma dr \simeq 2 \pi r^2 \Sigma$
at $r \ll r_0$.
The disk accretion rate is $\dot{M}_{\rm g} \sim 3 \pi \Sigma \nu_{\rm acc} \sim 3 \pi \alpha_{\rm acc} (h/r)^{2} r^2 \Omega$ at $r \ll r_0$.
For the disk accretion, $\alpha_{\rm acc}$ must be used.
Observationally, it is inferred that $\alpha_{\rm acc} \sim 10^{-3}-10^{-2}$ as below.
We use the disk accretion rate $\dot{M}_{\rm g}$ as a parameter and
$\Sigma$ is calculated by $\dot{M}_{\rm g}$ as
\begin{equation}
\Sigma \sim \frac{\dot{M}_{\rm g}}{3 \pi \nu_{\rm acc}}
\sim \frac{\dot{M}_{\rm g}}{3 \pi \alpha_{\rm acc} (h/r)^{2} r^2 \Omega}.
\label{eq:disk_acc}
\end{equation}
In the paper, the disk aspect ratio is given simply by $h/r \simeq 0.03 (r/1\,{\rm au})^{1/4}$.
Assuming the self-similar solution, $\Sigma \propto r^{-1} \exp(-r/r_0)$,
the total disk mass is $\int^\infty 2\pi \Sigma r dr = 2\pi r_0^2 \Sigma_0$.
Then, the global disk depletion timescale is given by
\begin{eqnarray}
t_{\rm dep} & \sim & \frac{2\pi \Sigma_0 r_0^2}{\dot{M}_{\rm g}} 
\sim \frac{2\pi \Sigma_0 r_0^2}{3 \pi \Sigma_0 \nu_{\rm acc,0}} 
= \frac{2r_0^2}{3\nu_{\rm acc,0}} 
= \frac{2}{3\Omega_0} \alpha_{\rm acc}^{-1} \left(\frac{h_0}{r_0}\right)^{-2} \label{eq:tdep1} \\
  & \simeq & 3 \times 10^6 \left(\frac{\alpha_{\rm acc}}{3 \times 10^{-3}} \right)^{-1}\left(\frac{(h/r)_{\rm 1au}}{0.03} \right)^{-2}
  \left(\frac{r_0}{100\,{\rm au}} \right)\; {\rm yrs}.
  \label{eq:tdep2}
\end{eqnarray} 
In the last equation, we used $h/r \propto r^{1/4}$.
Because observation of IR excess shows $t_{\rm dep}\sim \; {\rm a \,few} \times 10^6{\rm \, yr}$
and $r_0 \sim O(10^2)\,{\rm au}$ may be a reasonable value \citep[e.g.,][]{Williams11}, 
$\alpha_{\rm acc}$ is estimated to be $\sim 10^{-3}-10^{-2}$.
The value of $\alpha_{\rm acc}$ is constrained by the observation,
while $\alpha_{\rm vis}$ is theoretically estimated.

As we already pointed out, it is recently 
suggested that MRI in the disks is usually weak due to non-ideal MHD effects
and that wind-driven accretion is responsible for the global disk depletion.
In that case, $\alpha_{\rm vis}$ by turbulent viscous diffusion is much smaller than 
the effective $\alpha_{\rm acc}$ by the disk wind \citep{Armitage13}.
\citet{Hasegawa17} theoretically evaluated that $\alpha_{\rm vis} \sim 0.1 \, \alpha_{\rm acc}$.\footnote{Note that $\alpha_{\rm vis}$ in the present paper corresponds to
$\alpha_{\rm SS}$ in \citet{Hasegawa17} and to $\alpha$ in \citet{Armitage13};
$\alpha_{\rm acc}$ here corresponds to
$\alpha_{\rm SS,eff}$ in \citet{Hasegawa17} and 
to $\alpha + (4r/(3\sqrt{\pi} h)) \mid \overline{W_{z\phi}} \mid$ in \citet{Armitage13}.}
In the case of wind-driven accretion, the surface density gradient may be
less steep than Eq.~(\ref{eq:SSS}),
but in our discussions here, 
we just use $\Sigma \sim \dot{M}_{\rm g}/({\rm sevaral}\times \nu_{\rm acc})$
and $t_{\rm dep} \sim ({\rm disk \, mass})/\dot{M}_{\rm g}$ 
at $r \sim 1-10$ au, which would not be changed.

From Eqs.~(\ref{eq:cl2}) and (\ref{eq:tdep1}), 
\begin{eqnarray}
\frac{t_{\rm mig2,cl}}{t_{\rm dep}} & \sim &
\frac{\Omega_0}{\Omega} \left(\frac{h/r}{h_0/r_0}\right)^{-2} 
\left(1 +\frac{M_{\rm p}}{2 \pi \Sigma r^2} \right) \sim \left(\frac{r}{r_0}\right)
 \left(1 +\frac{M_{\rm p}}{2 \pi \Sigma r^2} \right). 
 \label{eq:mig_dep_ratio}
\end{eqnarray} 
Because usually $r \ll r_0$, it is predicted that
$t_{\rm mig2,cl}/t_{\rm dep} \ll 1$.
The only exception is 
an extremely depleted phase with 
$2 \pi \Sigma r^2/M_{\rm p} < r/r_0 \ll 1$.
Its contribution to the total migration is very limited.
Therefore, in the classical model, 
it is predicted that gas giants
usually undergo significant type II migration \citep[e.g.,][]{IL08, Hasegawa13,Bitsch15b}.
Note that $\alpha_{\rm acc}$ cancels out in Eq.~(\ref{eq:mig_dep_ratio})
and the conclusion here is independent of $\alpha_{\rm acc}$ and $\alpha_{\rm vis}$.
We discuss another possibility to avoid significant type II migration 
in the next subsection.

\subsection{Competition between migration and growth}

If the inner disk mass is reduced by external photoevaporation (AMB) 
or by accretion onto the planet \citep{TT16}, 
the planet-dominated migration starts earlier
and the total migration distance becomes smaller.
Planet-dominated migration is also realized,
if the growth due to gas accretion always dominates over the migration.
This is another possibility to avoid significant type II migration.
To examine this possibility, the
growth timescale ($t_{\rm grow}$) and migration timescale ($t_{\rm mig2}$)
are compared.
We will show that RV data cannot be reproduced as long as 
the classical migration formula is used.  

The critical core mass, beyond which hydrodynamic pressure no longer supports
the gas envelope against the planetary gravity,
is given by \citep[e.g.,][]{Ikoma00}
\begin{equation}
M_{\rm crit} \simeq
10 \left[\left( \frac{\dot{M}_{\rm c}}{10^{-6}M_{\oplus}/ {\rm yr}}\right)
\left(\frac{\kappa}{1\,{\rm cm^{2}/g}}\right) \right]^{k1}
M_{\oplus},
\label{eq:crit_core_mass}
\end{equation}
where $\dot{M}_{\rm c}$ is a solid core accretion rate and $k1 \sim 0.2-0.3$.
Growth of a gas giant through gas accretion onto the planet
is regulated by Kelvin-Helmholtz (KH) quasi-static contraction of the gas envelope
until the growth becomes so rapid that the supply of disk gas regulates the growth.
The mass growth rate in the KH phase is given by 
\citep[e.g.,][]{Ikoma00,IkomaGenda06}
\begin{equation}
\frac{dM_{\rm p,KH}}{dt} \simeq \frac{M_{\rm p}}{\tau_{\rm KH}},
\label{eq:mgsdot}
\end{equation}
where 
\begin{equation}
\tau_{\rm KH} \sim 10^3
\left(\frac{M_{\rm p}}{100\, M_{\oplus}}\right)^{-k2} \left(\frac{\kappa}{1\,{\rm cm^{2}/g}}\right)\, {\rm yrs},
\label{eq:tau_KH}
\end{equation}
with $k2 \sim 2.5-3.5$.
For simplicity, we adopt $k2 = 3$ and $\kappa=1\,{\rm cm^{2}/g}$ according to IL.
AMB solved 1D internal structure, but they also obtained the solution
that can be fitted by a similar power-law as Eq.~(\ref{eq:tau_KH}) with slightly smaller values of $k2$.
Since gas accretion onto the planet is mostly regulated by
the supply from the disk as we will show below, the difference does not affect the results here.
With $k2 = 3$ and $\kappa=1\,{\rm cm^2/g}$,
\begin{equation}
\frac{dM_{\rm p,KH}}{dt} \simeq 3 \times 10^{-7}
\left(\frac{M_{\rm p}}{100\, M_{\oplus}}\right)^{4}  \, M_\odot/{\rm yr}.
\label{eq:mgsdot2}
\end{equation}
For $M_{\rm p} \ga 50 M_\oplus$,
the above accretion rate exceeds the typical disk gas accretion rate around CTTSs ($\sim 10^{-8}M_\odot/{\rm yr}$).

IL assumed that 
the gas accretion rate onto the planet vanishes
after the gap opening in the protoplanetary disk.
The gap is opened when 
both the thermal condition,
\begin{equation}
h < r_{\rm H}, 
\label{eq:thermal}
\end{equation}
where $r_{\rm H}$ is the Hill radius of the planet
($r_{\rm H}=(M_{\rm p}/3 M_*)^{1/3} r$),  
and the viscous condition, 
\begin{equation}
\frac{M_{\rm p}}{M_*} > \frac{40 \nu_{\rm vis}}{r^2 \Omega},
\label{eq:viscous}
\end{equation} 
are satisfied 
\citep{LinPapaloizou93, Crida06}, 
where $M_*$ is the mass of the host star and 
$\nu_{\rm vis} = \alpha_{\rm vis}(h/r)^2 r^2 \Omega$ is the turbulent viscosity.
Because the thermal condition generally requires a larger planetary mass,
\citet{IL13} imposed a gas supply limit to the planet due to the gap opening as
$\dot{M}_{\rm p,gap}\sim \dot{M}_{\rm g} \exp(- M_{\rm p}/M_{\rm p,th})$, 
where $M_{\rm p,th} \sim 120(r/1{\rm au})^{3/4} M_\oplus$ that is 
equivalent to $r_{\rm H}\sim 2h$.
That is, the IL formula for gas accretion rate onto a planet is 
\begin{equation}
\frac{dM_{\rm p}}{dt} \simeq \min \left[\frac{dM_{\rm p,KH}}{dt}, \dot{M}_{\rm g} \exp\left(-\frac{M_{\rm p}}{M_{\rm p,th}}\right) \right].
\label{eq:mpdot1}
\end{equation}

\begin{figure}[htbp]
  \centering
  \includegraphics[width=150mm, angle = 0]{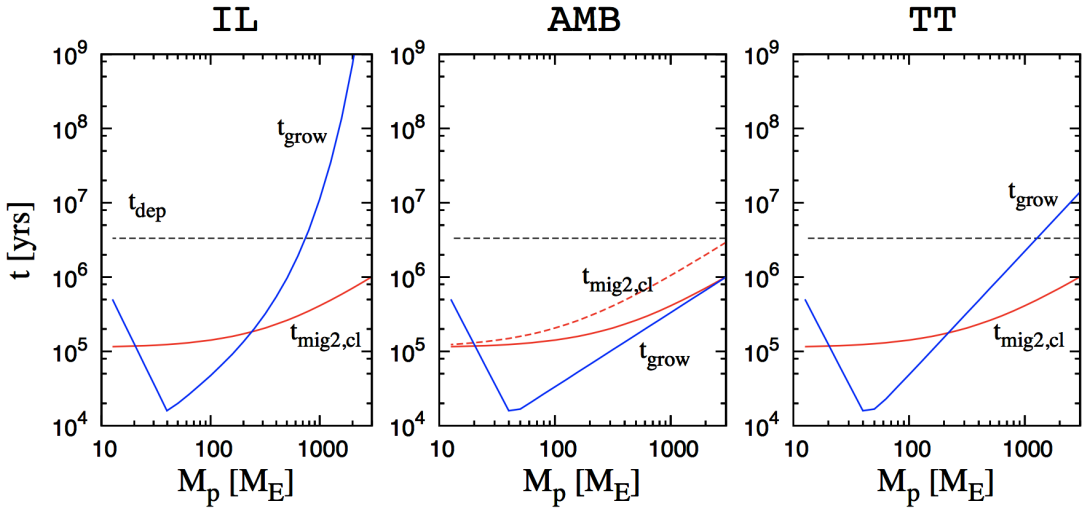}
  \caption{
Timescale of classical type II migration ($t_{\rm mig2,cl}$) 
is compared with that of 
growth through gas accretion of a planet ($t_{\rm grow}$) at 3 au,
as a function of the planet mass.
The left, middle, and right panels show $t_{\rm grow}$
by the IL, AMB, and TT growth formulas, respectively. 
The global depletion timescale $t_{\rm dep}=3 \times 10^{6}$ yrs
derived from $\alpha_{\rm acc}=3\times 10^{-3}$ and $r_0=100$ au
is also plotted.
The disk accretion rate $\dot{M}_{\rm g} = 10^{-8}M_\odot/{\rm yr}$ is assumed.
}
  \label{fig:tmig3cl}
\end{figure}

In the left panel of Fig.~\ref{fig:tmig3cl}, we plot $t_{\rm mig2,cl}$ given by Eq.~(\ref{eq:cl2})
and the growth timescale defined by $t_{\rm grow}=M_{\rm p}/(dM_{\rm p}/dt)$ with the IL formula
(Eq.~\ref{eq:mpdot1}).
The gradual increase in $t_{\rm mig2,cl}$ for $M_{\rm p} \ga M_{\rm J} \simeq 300 M_\oplus$ is 
the effect of the planet-dominated regime.
For $M_{\rm p} < 40 M_\oplus$, the Kelvin-Helmholtz contraction is a bottleneck.
Otherwise, the supply limit is more important. 
In the range of $M_{\rm p} \simeq 20M_\oplus-200M_\oplus$, 
$t_{\rm grow} < t_{\rm mig2,cl}$ and
the planet does not actually start migration until it grows up to $M_{\rm p} \simeq 200M_\oplus$.
Because the growth timescale at $M_{\rm p} \simeq 200M_\oplus$ is a few $\times 10^5$ yrs
and it is still 10 times shorter than $t_{\rm dep}$, 
the planet should further grow through accretion of gas 
(except in the case where the core of the gas giant is formed after the disk gas is highly depleted)
to enter the migration-dominated phase 
($t_{\rm grow} > t_{\rm mig2,cl}$)
and eventually undergo significant migration.

\begin{figure}[htbp]
\centering
\includegraphics[width=160mm, angle = 0]{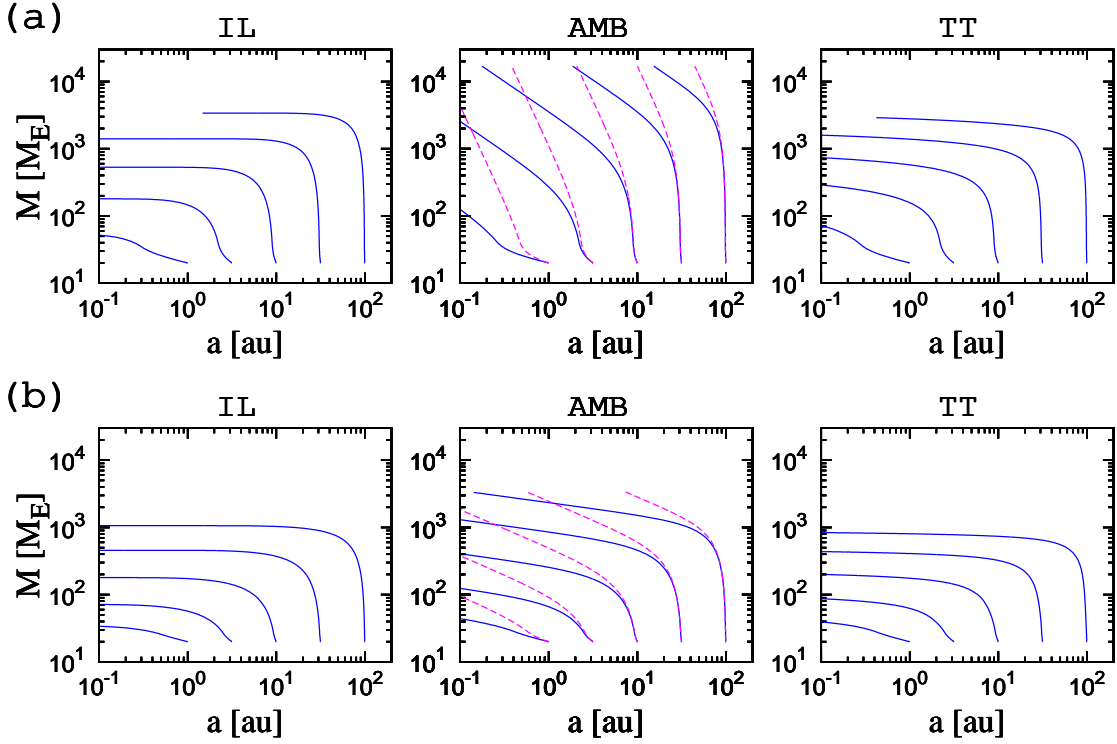} \vspace{2cm} 
  \caption{
Theoretical predictions of migration and growth of planets
due to gas accretion onto the planets with
initial semimajor axes 1, 3, 10, 30 and 100 au.
In the upper panels (a),
the left, middle, and right panels are the results with the growth formulas by the IL, AMB, and TT, respectively. 
In the middle panels, the dashed lines are the results with a planet-dominated
factor by a factor of $\pi$ larger than the nominal one in Eq.~(\ref{eq:cl2}).
In the lower panels (b), the growth rates are reduced by a factor of 5
(see the discussion in section 2.3).
In all cases, the initial masses are $20 M_\oplus$, 
$\alpha_{\rm acc} = 3 \times 10^{-3}$, $\alpha_{\rm vis} = 3 \times 10^{-4}$, $t_{\rm dep}=3 \times 10^6 \,{\rm yrs}$, and $\dot{M}_{\rm g} = 10^{-8}M_\odot/{\rm yr}$.
}
  \label{fig:evol}
\end{figure}

The mass and semimajor axis evolution of planets with the classical type II
migration formula (Eq.~\ref{eq:cl2}) and the growth formula
given by Eq.~(\ref{eq:mpdot1}) is shown in the left panel of Fig.~\ref{fig:evol}a.
As shown in the left panel of Fig.~\ref{fig:tmig3cl},
after $M_{\rm p}$ exceeds $\sim M_{\rm p,th}$, the growth rate decays
and the migration dominates ($t_{\rm mig2} \ll t_{\rm grow}, t_{\rm dep}$). 
Since $M_{\rm p,th}\sim 120(r/1{\rm au})^{3/4} M_\oplus$,
the transition mass for migration-dominance increases with $r$.
In this plot, all the planets 
migrate to the innermost regions of $a < 0.1\,{\rm au}$
except in the case starting at $100\,{\rm au}$.
Therefore, IL needed to introduce an artificial slowdown factor
in order to retain gas giants in the outer regions.

The timescales and evolution paths
with the AMB growth formula are shown in 
the middle panels of Figs.~\ref{fig:tmig3cl} and \ref{fig:evol}a.
AMB assumed
that the gas accretion rate onto the planet is always equal to $\dot{M}_{\rm g} = 3 \pi \Sigma \nu_{\rm acc}$
that is not impeded at all by the gap.
After the second term in the numerator in Eq.~(\ref{eq:cl}) becomes dominant
and the migration is transferred to planet-dominated one,
$t_{\rm grow} = M_{\rm p}/\dot{M}_{\rm p} \simeq M_{\rm p}/\dot{M}_{\rm g} \simeq t_{\rm mig2,cl}$,
that is, $d \log M/d \log a \simeq - 1$
(solid curves in the middle panel of Fig.~\ref{fig:tmig3cl}).
Note that AMB adopted the planet-dominated term
that is a factor $\pi$ larger than that  
in Eq.~(\ref{eq:cl2}) to obtain $d \log M/d \log a \simeq - \pi$ in their simulations
(the dashed curves in the middle panel of Fig.~\ref{fig:evol}).\footnote{
The value of $d \log M/d \log a$ is slightly different from $- \pi$ in 
the prescription of \citet{Mordasini2012}.   
}
In the AMB growth formula, 
$t_{\rm grow}$ is always $\la t_{\rm mig2,cl}$ and growth is dominant,
as shown in the middle panel of Fig.~\ref{fig:evol}a.

The AMB growth formula assumes the extreme limit of fast planetary growth rate.
Because all the disk accretion flow is captured by the planet in their formula
and most of disk mass initially resides in outer regions,
the planet's mass inevitably becomes comparable to
the initial total disk mass.
It would be much more than $10 M_{\rm J}$, which is
the upper limit of exoplanet mass inferred from RV observations.
To truncate the planetary growth,  
strong external photoevaporation was introduced in the AMB formula.
But, because truncation of planet growth due to gas accretion
is equivalent to truncation of migration, 
hot jupiters with $a \la 0.1$ au are scarcely formed \citep{AMB13}. 
Furthermore, it is not clear if external photoevaporation
is responsible for disk gas depletion in most of disks.

While the AMB formula assumed fully unimpeded gas accretion flow onto the planet, 
hydrodynamical simulations \citep[e.g.,][]{D'Angelo03, Machida10}
show that the accretion flux from the protoplanetary disk to the planet decreases 
as the planet mass increases and the gap becomes deeper (also see the discussion in section 2.3).\footnote{Note that the disk can become
eccentric by the planetary perturbations, when $M_{\rm p} \ga 5 M_{\rm J}$ \citep{Kley06}, which
may suppress the decay of the accretion flux onto the planet, excite the eccentricity of
the planetary orbit, and slow down the migration due to the supersonic effect
on planet-disk interaction \citep[e.g.,][]{Papaloizou00}.
Such effects will be left for future work.}
Small-mass embedded planets cannot capture
all the disk gas flow either.

TT proposed that
the gas accretion rate onto the planet is 
determined by the local gas surface density in the accretion band.
They derived the maximum accretion rate
of a relatively small planet that is embedded in the disk gas as 
\begin{eqnarray}
\dot{M}_{\rm p,nogap} 
& \simeq & 0.29 \left(\frac{h}{r}\right)^{-2}
\left(\frac{M_{\rm p}}{M_*}\right)^{4/3}\Sigma r^2\Omega 
\simeq \frac{0.29}{3\pi} \left(\frac{h}{r}\right)^{-4}
\left(\frac{M_{\rm p}}{M_*}\right)^{4/3}
\left(\frac{\dot{M}_{\rm g}}{\alpha_{\rm acc}}\right)  \label{eq:mgnogap}\\
 & \simeq & 3.8 \, \alpha_{\rm acc}^{-1} 
 \left(\frac{(h/r)_{\rm 1au}}{0.03} \right)^{-4}
 \left(\frac{r}{\rm 1\,au}\right)^{-1}
\left(\frac{M_{\rm p}}{M_{\rm J}}\right)^{4/3} \dot{M}_{\rm g}.
\end{eqnarray} 
For $M_{\rm p}$ significantly smaller than $M_{\rm J}$, $\dot{M}_{\rm p,nogap} < \dot{M}_{\rm g}$.
The planet accretion rate $\dot{M}_{\rm p,nogap}$ increases with $M_{\rm p}$. 
TT found that when the planet becomes massive enough to open up a gap in the disk,
the accretion rate is significantly reduced from the above formula.
The growth timescale and evolution paths by the TT formula 
are shown in
the right panels of Figs.~\ref{fig:tmig3cl} and \ref{fig:evol}.
To explain the TT formula, we summarize
the recent understanding on the gap opening in the next subsection.

\subsection{Gap opening}

The accretion flux onto the planet in a gap is
obtained by replacing the unperturbed surface density $\Sigma$ by 
the local surface density at the accretion band (TT).
Recent hydrodynamical simulations \citep{Duffell13, Fung14, Kanagawa15}
show that the gap is wide enough that the local surface density at the accretion band
is approximated by the surface density at the bottom of the gap ($\Sigma_{\rm min}$),
and $\Sigma_{\rm min}$ is given by
\begin{equation}
\frac{\Sigma_{\rm min}}{\Sigma} \simeq (1 + 0.04 K)^{-1},
\label{eq:gap_0ep}
\end{equation} 
where 
\begin{equation}
K = \left(\frac{M_{\rm p}}{M_*}\right)^2 \left(\frac{h}{r}\right)^{-5} \alpha_{\rm vis}^{-1}.
\label{eq:K}
\end{equation} 
This gap depth factor $K$ is related to the theoretical arguments of
the thermal and viscous conditions for the gap opening.
Since the gap opening is affected by turbulent viscous diffusion,
$\alpha_{\rm vis}$ is used here.
Equations~(\ref{eq:thermal}) and (\ref{eq:viscous}) are equivalent to
\begin{eqnarray}
1 < K_{\rm t} & \simeq & \frac{1}{3} \left(\frac{M_{\rm p}}{M_*}\right) \left(\frac{h}{r}\right)^{-3} 
\simeq \frac{1}{3} \left(\frac{M_{\rm p}}{M_*}\right) \left(\frac{h}{r}\right)^{-3} 
\\
1 < K_\nu & \simeq & \frac{1}{40} \left(\frac{M_{\rm p}}{M_*}\right) \left(\frac{h}{r}\right)^{-2} \alpha_{\rm vis}^{-1}.
\end{eqnarray} 
The gap depth factor $K$ is proportional to $K_{\rm t} K_\nu$.

Accordingly, TT proposed that $f_{\rm local} \dot{M}_{\rm g}$ limits the growth rate,
where $f_{\rm local}$ is the accretion efficiency given by
\footnote{We use the numerical factor 0.04 for $K$ by \citet{Kanagawa18}
rather than 0.034 by TT.} 
\begin{equation}
f_{\rm local} = \left(\frac{\dot{M}_{\rm p,nogap}}{1+ 0.04 K}\right) \dot{M}_{\rm g}^{-1} 
\simeq \frac{0.031 \left(\frac{h}{r}\right)^{-4} \left(\frac{M_{\rm p}}{M_*}\right)^{4/3}
\alpha_{\rm acc}^{-1}}{1+ 0.04 K}
\simeq \frac{3 \left(\frac{(h/r)_{\rm 1au}}{0.03} \right)^{-4} 
\left(\frac{r}{\rm 1au} \right)^{-1}
\left(\frac{M_{\rm p}}{M_{\rm J}}\right)^{4/3}
\alpha_{\rm acc}^{-1}}{1+ 0.04 K}.
\label{eq:f_local}
\end{equation}
We add the Kelvin-Helmholtz contraction term and
here the TT formula is defined by 
\begin{equation}
\frac{dM_{\rm p}}{dt} \simeq 
\min \left[\frac{dM_{\rm p,KH}}{dt}, \dot{M}_{\rm g}, f_{\rm local} \dot{M}_{\rm g}\right].
\label{eq:mpdot3}
\end{equation}
Before the gap opening ($0.04K \ll 1$), $f_{\rm local}$ increases with $M_{\rm p}$,
because the planetary gravity becomes stronger.
However, once the gap opening becomes important at $0.04K > 1$, that is,
\begin{eqnarray}
M_{\rm p} > M_{\rm p,crit} & \equiv & 5 \alpha_{\rm vis}^{1/2}\left(\frac{h}{r}\right)^{5/2} M_*\\
 & \simeq & 0.014 \left(\frac{\alpha_{\rm vis}}{3 \times 10^{-4}}\right)^{1/2}
\left(\frac{(h/r)_{\rm 1au}}{0.03} \right)^{5/2}
\left(\frac{r}{1\,{\rm au}} \right)^{5/8} M_{\rm J},
\label{eq:f_local_crit}
\end{eqnarray}
$f_{\rm local}$ decreases as $M_{\rm p}$ increases, because of the surface density depletion 
in the accretion band associated by the gap opening.
In this case, $f_{\rm local}$ is approximated by
\begin{equation}
f_{\rm local} \simeq 0.78 \frac{\alpha_{\rm vis}}{\alpha_{\rm acc}}\frac{h}{r}\left(\frac{M_{\rm p}}{M_*}\right)^{-2/3}
\simeq 
\left(\frac{\alpha_{\rm vis}/\alpha_{\rm acc}}{0.1}\right)
\left(\frac{(h/r)_{\rm 1au}}{0.03} \right)
\left(\frac{r}{1\,{\rm au}} \right)^{1/4}
  \left(\frac{M_{\rm p}}{0.1 M_{\rm J}}\right)^{-2/3}.
\label{eq:f_local2}
\end{equation}
As we pointed out, the gap is deeper for lower $\alpha_{\rm vis}$,
while the gas accretion rate through the disk is proportional to $\alpha_{\rm acc}$.
Accordingly, the accretion efficiency $f_{\rm local}$ 
is proportional to $\alpha_{\rm vis}/\alpha_{\rm acc}$.
  
While TT implicitly assumed $\alpha_{\rm vis} = \alpha_{\rm acc}$,
the local supply limit is very important 
in the case of $\alpha_{\rm vis} \ll \alpha_{\rm acc}$,
when the gap is deep\footnote{Because TT only discussed the case of $\alpha_{\rm vis} = \alpha_{\rm acc}$,
the availability of Eq.~(\ref{eq:mpdot3}) must be examined in the case of $\alpha_{\rm vis} \ll \alpha_{\rm acc}$.}.
For $\alpha_{\rm vis}/\alpha_{\rm acc}\sim 1/10$, which is consistent with
analysis of the disk wind and MRI by \citet{Hasegawa17},
the local supply limit becomes already important ($f_{\rm local} < 1$)
when $M_{\rm p} \ga 0.15 M_{\rm J}$ at 3 au.


The growth timescale by the TT formula is shown in
the right panel of Fig.~\ref{fig:tmig3cl}, which is more similar to
the IL formula than to the AMB one: $t_{\rm grow} > t_{\rm mig2,cl}$
for $M_{\rm p} \ga 100 M_\oplus$ at 3 au.
It suggests that a planet formed at $\sim 3$ au undergoes significant
migration when $M_{\rm p}$ becomes $\ga 100 M_\oplus$,
which is clearly shown in the right panel of Fig.~\ref{fig:evol}a.

Previous models implicitly assumed that most of the disk gas entering
the planetary Hill sphere is accreted by the planet, based on the results of
isothermal hydrodynamical simulations. 
\citet{Kurokawa18} and Lambrechts et al. (2018, in prep.) performed non-isothermal
hydrodynamical simulations for an embedded planet to find that 
gas flow is prevented from entering deep regions of the planetary atmosphere
by the non-isothermal effect. 
\citet{Szulagyi16} and \citet{Szulagyi17} showed 
that the gas flow onto a circumplanetary disk around a jupiter-mass planet 
is meridional circulation and a significant fraction of infalling gas
is lost from the outer part of the disk. 
These results could suggest a lower accretion rate in the non-isothermal case
than in the isothermal case.
Simulations with low resolution near the Hill radius
also tend to show a higher accretion rate \citep{Tanigawa02}.
Considering these possibilities, we tested the cases with
the disk gas supply rates reduced by a factor of 5 in Fig.~\ref{fig:evol}b.
Because the growth is slowed down, the migration becomes more dominant
and type II migration becomes more significant, which makes the problem more serious.
Even with the AMB formula, 
$t_{\rm mig2,cl}$ becomes shorter than $t_{\rm grow}$ and
all the gas giants significantly migrate. 

As we have shown in section 2.2, as long as the classical formula is adopted,
$t_{\rm mig2,cl} < t_{\rm grow}$ or,
at most, $t_{\rm mig2,cl} \sim t_{\rm grow}$.
In other words, type II migration is significant
except for the extreme case where  
all the gas flow through the protoplanetary disk is assumed to be accreted by the planet
as in the AMB formula.
The most recent TT formula shows that
only a fraction of the gas accreting through the disk is accreted by the planet
($f_{\rm local} < 1$) for $M_{\rm p} \ga 0.1(r/{\rm 1 au})^{3/8} M_{\rm J}$ (see Eq.~\ref{eq:f_local2}).


However, with the new formula of type II migration has been proposed by
\citet{Kanagawa18}, the serious problem of type II migration
can be solved.
TT suggested that the disk mass interior to the planet 
is reduced by gas accretion onto the planet by a factor of $(1 + f_{\rm local})$.
If it is in the planet-dominated regime, type II migration 
is slowed down by the same factor.
We will show that type II migration is sufficiently slowed down by
the new formula even without taking this effect into account.

\section{New model of type II migration}

Recent hydrodynamic simulations \citep[e.g.,][]{Duffell14,Durman15} argued that type II migration of a gap-opening gas giant is {\it not} tied to disk gas accretion.
By carrying out 2D hydrodynamic simulations in a broad range of parameters, \citet{Kanagawa18} confirmed this argument and constructed an empirical formula for the migration speed of the gap-opening planet. 
For $\alpha_{\rm vis} \sim \alpha_{\rm acc}$, the new formula is consistent with other simulations 
in the literatures.
On the other hand, it shows a significant slowdown of type II migration 
for $\alpha_{\rm vis} \ll \alpha_{\rm acc}$
and for gas giants with $M_{\rm p} \ga M_{\rm J}$.
The new formula needs to be confirmed by detailed follow-up simulations.
Here we investigate its impact on evolution of gas giants, assuming that it is correct.  

\citet{Kanagawa18}'s formula indicates that the migration of the gap-opening planet is decoupled from the disk gas accretion and it is determined by the disk-planet interaction with the gas at the bottom of the gap.
According to their empirical formula for the migration speed of the gap-opening planet,
the type II migration timescale is nothing other than the (corotation-torque saturated)  
type I migration timescale with the reduced gas surface density ($\Sigma_{\rm min}$) in the gap,
\begin{equation}
t_{\rm mig2} \simeq \frac{\Sigma}{\Sigma_{\rm min}} t_{\rm mig1},
\label{eq:tmig2_new}
\end{equation} 
where $t_{\rm mig1}$ is the type I migration timescale given by
\begin{equation}
t_{\rm mig1} \simeq \frac{1}{2c}\left(\frac{M_{\rm p}}{M_*}\right)^{-1} \left(\frac{\Sigma r^2}{M_*}\right)^{-1} \left(\frac{h}{r}\right)^{2} \Omega^{-1}.
\label{eq:tmig1K}
\end{equation} 
\citet{Kanagawa18} showed that $c\sim 1-3$ and
the above $t_{\rm mig1}$ is consistent with the isothermal formula for type I migration by 
\citet{Tanaka02}. 
For planets that undergo type I migration, $\Sigma_{\rm min} \sim \Sigma$.
Planet masses that undergo type II migration would be well above
the masses affected by non-saturated corotation torque \citep[e.g.,][]{Paardekooper11}.
Thus, with their formula, type I and type II migrations are continuous and
the uncertainty on switching from type I migration to type II migration, 
which has long been one of the big ambiguities in population synthesis simulations, is resolved.  

\begin{figure}[htbp]
  \centering
  \includegraphics[width=100mm, angle = 0]{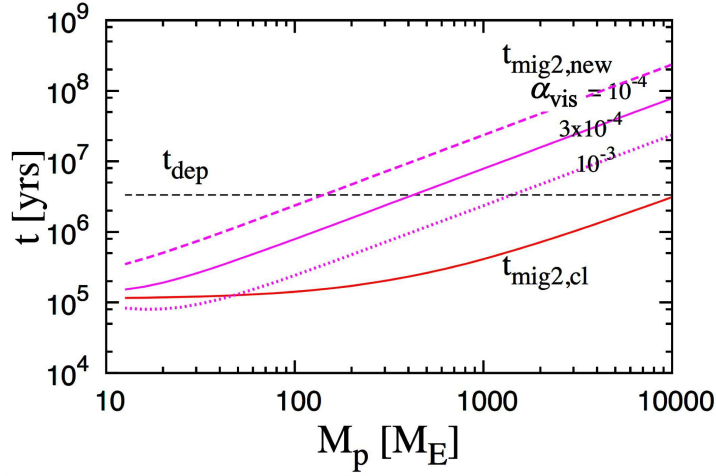}
  \caption{
Theoretical predictions of type II migration timescales at 3 au
with $\alpha_{\rm acc}=3\times 10^{-3}$,
as a function of the planet mass ($M_{\rm p}$).
The classical formula is represented by the red curve
and the new formula is represented by the magenta curves
for $\alpha_{\rm vis}=10^{-4}, 3 \times 10^{-3}$ and $10^{-4}$.
The disk accretion rate, $\dot{M}_{\rm g} = 10^{-8}M_\odot/{\rm yr}$, is assumed.
The global depletion timescale $t_{\rm dep}=3 \times 10^{6}$ yrs
with  $\alpha_{\rm acc}=3\times 10^{-3}$ and $r_0=100$ au
is also plotted.
}
  \label{fig:tmignew}
\end{figure}

\begin{figure}[htbp]
  \centering 
 \includegraphics[width=150mm, angle = 0]{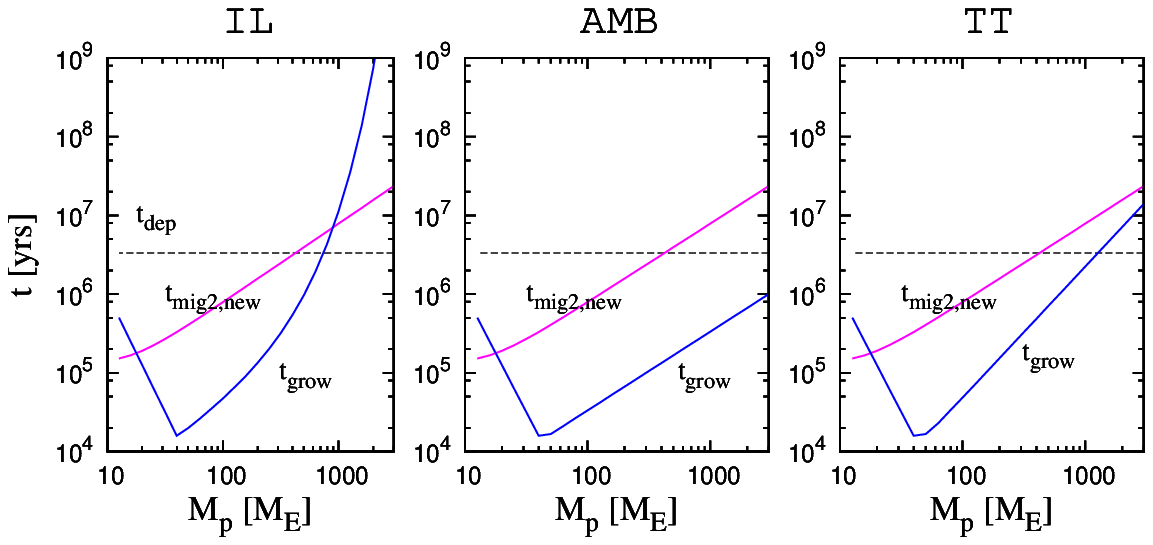} 
\vspace*{1cm} 
  \caption{
Same as Fig.~\ref{fig:tmig3cl} except
that the type II migration timescale is given by the new formula (Eq.~\ref{eq:tmig2new})
where $\alpha_{\rm vis} = 0.1 \times \alpha_{\rm acc} = 3\times 10^{-4}$ is used.}
  \label{fig:tmig3new}
\end{figure}

From Eqs.~(\ref{eq:gap_0ep}), (\ref{eq:K}), (\ref{eq:tmig2_new}), and (\ref{eq:tmig1K}),
the new formula of type II migration by \citet{Kanagawa18} is written as
\begin{eqnarray}
t_{\rm mig2,new} & \simeq & (1+ 0.04 K) \, t_{\rm mig1} \label{eq:tmig2new0}\\
  & \simeq & \frac{1}{2c}
\left[1+0.04 \left(\frac{M_{\rm p}}{M_*}\right)^2 \left(\frac{h}{r}\right)^{-5} \alpha_{\rm vis}^{-1}\right]
\left(\frac{M_{\rm p}}{M_*}\right)^{-1} \left(\frac{\Sigma r^2}{M_*}\right)^{-1} \left(\frac{h}{r}\right)^{2} \Omega^{-1}.
\label{eq:tmig2new}
\end{eqnarray} 
For $0.04 K \gg 1$ and $c \sim 2$,
\begin{eqnarray}
t_{\rm mig2,new} & \simeq & 
0.01 \alpha_{\rm vis}^{-1} \left(\frac{M_{\rm p}}{\Sigma r^2}\right) \left(\frac{h}{r}\right)^{-3} \Omega^{-1} 
\simeq 0.1 \left(\frac{\alpha_{\rm acc}}{\alpha_{\rm vis}}\right) 
 \left(\frac{M_{\rm p}}{\dot{M}_{\rm g}}\right) \left(\frac{h}{r}\right)^{-1} \\
  & \simeq & 
3 \times 10^6 \left(\frac{\alpha_{\rm vis}/\alpha_{\rm acc}}{1/10} \right)^{-1}
 \left(\frac{(h/r)_{\rm 1au}}{0.03} \right)^{-1} \nonumber\\
 & & \left(\frac{r}{1\,{\rm au}}\right)^{-1/4}
\left(\frac{M_{\rm p}}{M_{\rm J}}\right)
\left(\frac{\dot{M}_{\rm g}}{10^{-8}M_\odot/{\rm yr}}\right)^{-1}
  \; {\rm yrs},
\label{eq:tmig2newB}
\end{eqnarray} 
({\small \bf NOTE: the power index -1 was missing for the $\alpha$ and $\dot{M}_{\rm g}$ terms in Eq.(31) in the previous \& published versions})
where we have used $\Sigma \simeq \dot{M}_{\rm g} / 3\pi \alpha_{\rm acc} h^2 \Omega$.
In Fig.~\ref{fig:tmignew}, we plot $t_{\rm mig2,new}$ given by Eq.~(\ref{eq:tmig2newB})
as well as $t_{\rm mig2,cl}$.
With the classical formula, the migration timescale gradually increases with $M_{\rm p}$
for $M_{\rm p}\ga M_{\rm J}$, 
with the effect of the planet-dominated regime.
However, with the new formula, the migration timescale increases more rapidly
with $M_{\rm p}$ and is much larger than
that with the classical formula, in particular, for
smaller values of $\alpha_{\rm vis}$.
The left, middle, and right panels of Figure~\ref{fig:tmig3new} show that
the new migration timescale $t_{\rm mig2,new}$ 
is longer than the growth timescales with the IL, AMB, and TT formulas, respectively,
in the range of $t < t_{\rm dep}$ except in the initial Kelvin-Helmholtz contraction phase
with $M_{\rm p} < 20  M_\oplus$. 

\begin{figure}[htbp]
  \centering
\includegraphics[width=160mm, angle = 0]{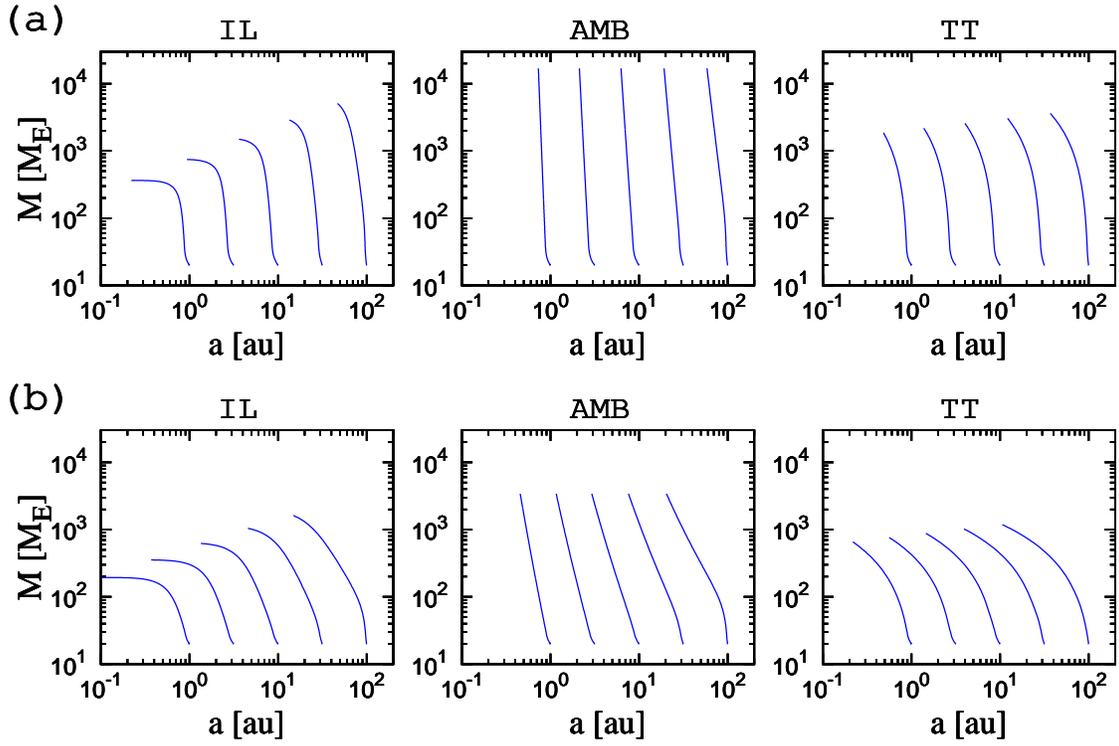}\vspace{2cm}
  \caption{
Same as Fig.~\ref{fig:evol} except
that the type II migration timescale is given by the new formula
with $\alpha_{\rm vis} = 0.1 \times \alpha_{\rm acc} = 3\times 10^{-4}$.
}
  \label{fig:evolnew}
\end{figure}

Figures~\ref{fig:evolnew}a shows the mass and semimajor axis evolution
of giant planets with the new type II migration formula.
The left, middle, and right panels show the results with the IL, AMB, and TT formulas
for planetary growth.
Type II migration is almost negligible for all the results.
Even if the reduction factor of 5 is applied for the disk gas supply rates,
most of gas giants survive against type II migration
(Fig.~\ref{fig:evolnew}b).

The TT formula uses 
the scaling law for the accretion band and
$\Sigma_{\rm min}$ given by Eq.~(\ref{eq:gap_0ep}) for the
accretion band gas surface density.
Because the new type II migration formula 
also uses $\Sigma_{\rm min}$ as in Eq.~(\ref{eq:tmig2_new}),
the TT formula could be more consistent with
the new formula.
However, the scaling law for the accretion band
is based on the local Hill's approximation ($(M_{\rm p}/M_*)^{1/3} \ll 1$)
and it is not accurate enough for massive planets.
Actually, TT pointed out that their predicted 
gas accretion rate onto the planet
is larger than that obtained by the previous hydrodynamical simulation results \citep{D'Angelo03, Machida10}
for $M_{\rm p} \sim M_{\rm J}$,
while they are consistent for $M_{\rm p} < M_{\rm J}$.
The previous hydrodynamical simulations
did not simulate the cases of $M_{\rm p} > M_{\rm J}$.
Figure 1 of TT may indicate  
that $f_{\rm local}$ decreases more rapidly with $M_{\rm p}$ 
for $M_{\rm p} > M_{\rm J}$ than their formula. 
In that case, the TT growth formula becomes similar to the IL one.
Thus, an actual gas accretion rate onto the planet still
includes uncertainty for $M_{\rm p} > M_{\rm J}$
and could be between the TT and the IL models
(the possibility of eccentric disks also exists, as pointed out
in the footnote 2 in section 2.2).
Detailed hydrodynamical simulations are needed
for $M_{\rm p} > M_{\rm J}$.

In Figs.~\ref{fig:evolnew}a and b,
we also plot the results with the IL and AMB formulas
in addition to those with the TT formula.
While the details are different, 
all the plots show slowdown of migration for planets with $M_{\rm p} \ga M_{\rm J}$.
Thus, the pile-up of gas giants beyond 1 au will be robustly reproduced
if the new type II migration formula is justified and $\alpha_{\rm vis} \ll \alpha_{\rm acc}$.

\section{Population synthesis simulations}

So far, our discussions started from cores with $M_{\rm p} = 20 M_\oplus$.
In this section, we show the results of planet population synthesis simulations
with the classical and new type II migration formulas.
Planet population synthesis calculates planetary growth and migration
from small planetesimals, combining planetesimal accretion,
gas accretion onto the planet, type I and II migrations, and planet-planet scattering.

Detailed prescriptions of the simulations are described in \citet{IL13},
except the new formula of type II migration.
We adopt the classical planetesimal accretion model.
The calculation with pebble accretion requires a model of detailed disk structure and its evolution, 
which is left for future work. 
We set embryos with an initial mass of $10^{20}$ g with orbital separations
of $\sim 10 \, r_{\rm H}$ of the classical isolation mass \citep{Kokubo_Ida98}---which 
means that the embryos are placed with a log uniform distribution---in a range of 0.05--20 au (planetesimal accretion is very slow beyond 20 au).
We use the self-similar disk model (Eq.~\ref{eq:SSS}).
The gas surface density at 10 au is distributed in a range of  
[0.1,10] times the minimum-mass solar nebula model \citep{Hayashi85}
with a log-normal function
for the solar-mass host stars.
The host star mass 
and the initial metallicity of the disk are distributed in ranges of
[0.8,1.25]$M_\odot$ with a log-normal function and 
[-0.2,0.2] dex with a normal function.
 
To highlight the effect of the new type II migration model, 
we use the simple IL model for the gas giant growth and 
type I migration timescale that is given by $30 \, t_{\rm mig1}$
where $t_{\rm mig1}$ is given by Eq.~(\ref{eq:tmig1K}).
In the new type II migration formula (Eq.~\ref{eq:tmig2new0}),
the reduction of the factor 30 is not applied.
This treatment is inconsistent with
the continuous transition from
the isothermal type I migration formula to the new type II migration formula.
However, for low-mass planets, if the isothermal type I migration formula
is directly applied, cores are removed and gas giants are scarcely formed
\citep[e.g.,][]{IL08,AMB09}.
Here we use the migration timescale
as $t_{\rm mig} = [1/(30 \, t_{\rm mig1}) + 1/t_{\rm mig2,new}]^{-1}$.
Recently, slowdown of the migration for low-mass planets 
from the isothermal formula has been
actively discussed.
For example, \citet{Paardekooper14} proposed that type I migration is
significantly slowed down by dynamical corotation torque in the case of
very low $\alpha_{\rm vis}$.
\citet{Ogihara17} argued that the disk wind decreases the gas surface density
in the inner disk regions and type I migration can be significantly slowed down.
In this paper, we do not go into details on the slowing down of type I migration. 

For a growth model of gas giants,
we use the IL growth formula (Eq.~\ref{eq:mpdot1}) rather than
that of TT (Eq.~\ref{eq:mpdot3}).
As we mentioned in section 3, the realistic model
could be between these.
With the TT growth formula, population synthesis simulations
produce gas giants with much more massive 
even in the close-in region, which is inconsistent with the observations,
unless strong external photoevaporation is applied, such as in the AMB formula.

\begin{figure}[htbp]
  \centering
  \includegraphics[width=100mm, angle = 0]{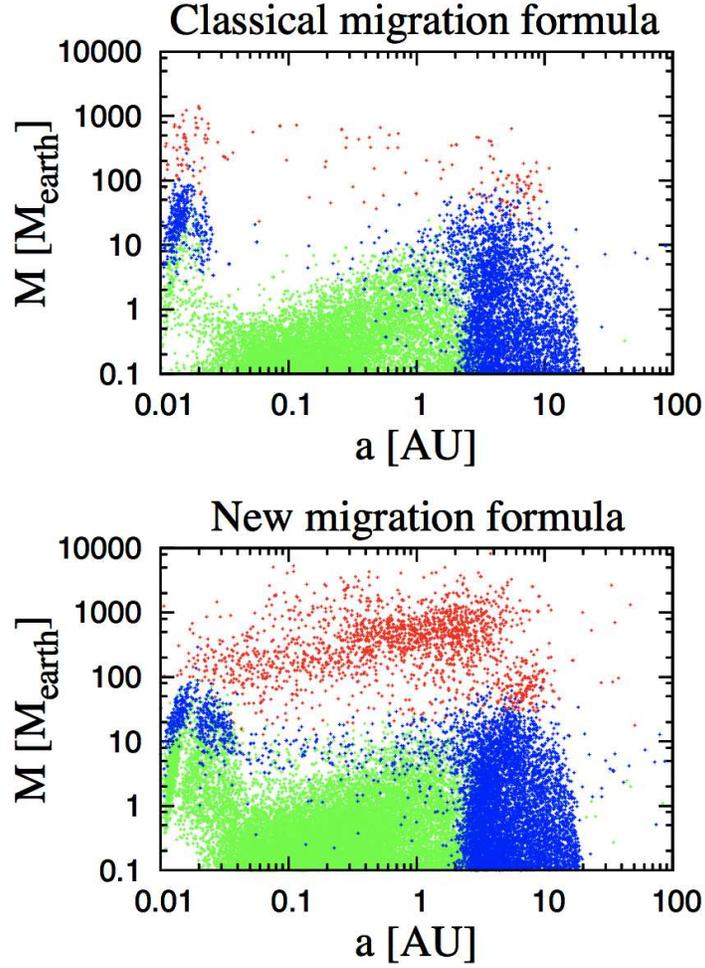}
  \caption{
Results of planet population synthesis simulations of 3000 systems
around solar-type stars.
In the upper panel, the classical type II migration formula without artificial reduction in speed
is used, while the new type II migration formula is used in the lower panel.
For the processes of planet formation other than the type II migration, we follow the prescriptions in \citet{IL13}.   
The red, blue, and green dots represent gas, icy, and rocky planets.  
We use $\alpha_{\rm acc}=3 \times 10^{-3}$ and $\alpha_{\rm vis} = 3\times 10^{-4}$.
}
  \label{fig:ma_pop}
\end{figure}

Figure~\ref{fig:ma_pop} shows the 
results of planet population synthesis simulations of 3000 systems
around solar-type stars.
In the upper panel, we use the classical type II migration formula
without any artificial reduction.
As discussed in section 3,
gas giants significantly migrate toward the host star
except a small fraction of the planets that were formed in the dissipating disks. 
In the results of Ida et al. (2013), the type II migration speed was artificially reduced
by a factor of 10 from the classical formula, as already pointed out.

In the lower panel, we use the new type II migration formula
with $\alpha_{\rm acc}=3 \times 10^{-3}$ and $\alpha_{\rm vis}=3 \times 10^{-4}$.
No artificial reduction is applied in the type II migration speed.
The concentration of gas giants at $\sim 0.5-5$ au is pronounced,
which is consistent with the RV observation data.
In the calculations, the snowline is simply set at $r_{\rm snow} = 2.7(M_*/M_\odot)\, {\rm au}$,
assuming optically thin disks.
Because the solid surface density is enhanced beyond the snowline,
cores large enough for runaway gas accretion emerge there before the disk gas is depleted.
With a small migration of both type I and II,
the final gas giants are concentrated at $\sim 0.5-5$ au.

Hot jupiters are formed by in situ gas accretion 
onto the cores that undergo type I migration
or by relatively faster type II migration of sub-jupiter-mass planets.
In our result, the gaseous planet mass is lower in the close-in region than 
in the outer region,
which is also consistent with the observations.

We also performed a simulation with the new type II migration formula
taking into account the effect of inner disk gas depletion due to accretion onto the planet,
proposed by TT.
The planet distribution is very similar to  
the lower panel of Figure~\ref{fig:ma_pop}
except for more depletion of massive giant planets at $a \la 0.5$ au,
which is more consistent with the RV data (Fig.~\ref{fig:Ma}).

\section{Summary}

The mass and semimajor axis distribution of gas giants in exoplanetary systems
is produced by complicated processes of 
competition between core formation and type I migration and
that between gas accretion onto the planets and type II migration
in evolving gas disks.
As explained in this paper it is very hard for theoretical predictions
with the classical type II migration model
to explain the RV data of gas giants in exoplanetary systems,
in particular, the pile-up of super-jupiter-mass planets beyond 1 au
and the broad semimajor axis distribution of sub-jupiter-mass planets
from 0.03 au to beyond 1 au.
Here, we have demonstrated that the newly proposed type II migration model
predicts the distribution of gas giant planets, consistent with the RV data.  

While the classical model assumed that planetary migration is tied to global gas accretion through the disk,
recent high-resolution simulations show that the gap is not clear enough
and the migration of the gap-opening planet is decoupled from the global disk gas accretion
\citep{Duffell13, Duffell14, Fung14, Kanagawa15}.
According to this new picture, 
a new type II migration formula was proposed by \citet{Kanagawa18} where they
argued that type II migration speed is nothing other than isothermal type I migration speed
with the reduced disk gas surface density in the gap.
We investigated the evolution of gas giants with the new type II migration formula
by comparing the migration timescale
with the growth timescale of gas giants and disk lifetime
and by performing population synthesis simulation.

We found that this new formula significantly slows down type II migration
for super-jupiter-mass planets that produce deep gaps.
If the alpha parameter for turbulent diffusion ($\alpha_{\rm vis}$) is small enough compared with
the effective alpha for global transfer of disk angular momentum and mass ($\alpha_{\rm acc}$),
which can be accounted for by the disk wind (wind-driven accretion), most of super-jupiter-mass gas planets
remain at $\sim 0.5-5$ au without any significant migration from their birthplaces.
Thus, the new type II migration formula may solve the problem
of the observed pile-up of gas giants at $a \ga 1$ au.

To reinforce our conclusion, more detailed follow-up fluid dynamical simulations 
are needed to test the new type II migration formula.
More detailed simulations are also needed
to investigate the growth rate of 
super-jupiter-mass planets regulated by gas supply across the gap 
and to evaluate the relevant values of $\alpha_{\rm acc}$ by the disk wind
and $\alpha_{\rm vis}$ due to turbulence from non-ideal MHD effects.

\vspace{1em} 
\noindent ACKNOWLEDGMENTS.  
We thank an anonymous referee for helpful comments
and Satoshi Okuzumi for detailed comments.
S.I. is supported by JSPS KAKENHI grant 15H02065.
A.J. is grateful to the Earth-Life Science Institute (ELSI) at Tokyo Institute of Technology for hosting his research visit during March 2018. A.J. further thanks the Swedish Research Council (grant 2014-5775), the Knut and Alice Wallenberg Foundation (grants 2012.0150, 2014.0017) and the European Research Council (ERC Consolidator Grant 724687-PLANETESYS) for research support.
K.D.K. was supported by the Polish National Science Centre MAESTRO 
grant DEC- 2012/06/A/ST9/00276
and by the JSPS Core-to-Core Program ``International Network of Planetary Sciences.''

\vspace{1em}
\noindent CORRESPONDENCE should be addressed to S. I. (ida@elsi.jp).

\clearpage

\bibliography{mig2_final} 

\begin{thebibliography}{}
\expandafter\ifx\csname natexlab\endcsname\relax\def\natexlab#1{#1}\fi

\bibitem[{{Alexander} \& {Pascucci}(2012)}]{Alexander2012}
{Alexander}, R.~D., \& {Pascucci}, I. 2012, \mnras, 422, L82

\bibitem[{{Alibert} {et~al.}(2013){Alibert}, {Carron}, {Fortier}, {Pfyffer},
  {Benz}, {Mordasini}, \& {Swoboda}}]{AMB13}
{Alibert}, Y., {Carron}, F., {Fortier}, A., {et~al.} 2013, \aap, 558, A109

\bibitem[{{Alibert} {et~al.}(2011){Alibert}, {Mordasini}, \& {Benz}}]{AMB11}
{Alibert}, Y., {Mordasini}, C., \& {Benz}, W. 2011, \aap, 526, A63

\bibitem[{{Armitage} {et~al.}(2013){Armitage}, {Simon}, \&
  {Martin}}]{Armitage13}
{Armitage}, P.~J., {Simon}, J.~B., \& {Martin}, R.~G. 2013, \apjl, 778, L14

\bibitem[{{Bai}(2017)}]{Bai17}
{Bai}, X.-N. 2017, \apj, 845, 75

\bibitem[{{Bai} {et~al.}(2016){Bai}, {Ye}, {Goodman}, \& {Yuan}}]{Bai16}
{Bai}, X.-N., {Ye}, J., {Goodman}, J., \& {Yuan}, F. 2016, \apj, 818, 152

\bibitem[{{Bitsch} {et~al.}(2015){Bitsch}, {Lambrechts}, \&
  {Johansen}}]{Bitsch15b}
{Bitsch}, B., {Lambrechts}, M., \& {Johansen}, A. 2015, \aap, 582, A112

\bibitem[{{Crida} {et~al.}(2006){Crida}, {Morbidelli}, \& {Masset}}]{Crida06}
{Crida}, A., {Morbidelli}, A., \& {Masset}, F. 2006, Icarus, 181, 587

\bibitem[{{Cumming} {et~al.}(2008){Cumming}, {Butler}, {Marcy}, {Vogt},
  {Wright}, \& {Fischer}}]{Cumming08}
{Cumming}, A., {Butler}, R.~P., {Marcy}, G.~W., {et~al.} 2008, \pasp, 120, 531

\bibitem[{{D'Angelo} {et~al.}(2010){D'Angelo}, {Durisen}, \&
  {Lissauer}}]{D'Angelo2010}
{D'Angelo}, G., {Durisen}, R.~H., \& {Lissauer}, J.~J. 2010, {Giant Planet
  Formation}, ed. S.~{Seager}, 319--346

\bibitem[{{D'Angelo} {et~al.}(2003){D'Angelo}, {Kley}, \&
  {Henning}}]{D'Angelo03}
{D'Angelo}, G., {Kley}, W., \& {Henning}, T. 2003, \apj, 586, 540

\bibitem[{{Duffell}(2015)}]{Duffell15}
{Duffell}, P.~C. 2015, \apjl, 807, L11

\bibitem[{{Duffell} {et~al.}(2014){Duffell}, {Haiman}, {MacFadyen}, {D'Orazio},
  \& {Farris}}]{Duffell14}
{Duffell}, P.~C., {Haiman}, Z., {MacFadyen}, A.~I., {D'Orazio}, D.~J., \&
  {Farris}, B.~D. 2014, \apjl, 792, L10

\bibitem[{{Duffell} \& {MacFadyen}(2013)}]{Duffell13}
{Duffell}, P.~C., \& {MacFadyen}, A.~I. 2013, \apj, 769, 41

\bibitem[{{D{\"u}rmann} \& {Kley}(2015)}]{Durman15}
{D{\"u}rmann}, C., \& {Kley}, W. 2015, \aap, 574, A52

\bibitem[{{Ercolano} \& {Rosotti}(2015)}]{Ercolano2015}
{Ercolano}, B., \& {Rosotti}, G. 2015, \mnras, 450, 3008

\bibitem[{{Fung} {et~al.}(2014){Fung}, {Shi}, \& {Chiang}}]{Fung14}
{Fung}, J., {Shi}, J.-M., \& {Chiang}, E. 2014, \apj, 782, 88

\bibitem[{{Hasegawa} \& {Ida}(2013)}]{Hasegawa13}
{Hasegawa}, Y., \& {Ida}, S. 2013, \apj, 774, 146

\bibitem[{{Hasegawa} {et~al.}(2017){Hasegawa}, {Okuzumi}, {Flock}, \&
  {Turner}}]{Hasegawa17}
{Hasegawa}, Y., {Okuzumi}, S., {Flock}, M., \& {Turner}, N.~J. 2017, \apj, 845,
  31

\bibitem[{{Hayashi} {et~al.}(1985){Hayashi}, {Nakazawa}, \&
  {Nakagawa}}]{Hayashi85}
{Hayashi}, C., {Nakazawa}, K., \& {Nakagawa}, Y. 1985, in Protostars and
  Planets II, ed. D.~C. {Black} \& M.~S. {Matthews}, 1100--1153

\bibitem[{{Ida} \& {Lin}(2004{\natexlab{a}})}]{IL04}
{Ida}, S., \& {Lin}, D.~N.~C. 2004{\natexlab{a}}, \apj, 604, 388

\bibitem[{{Ida} \& {Lin}(2004{\natexlab{b}})}]{IL04b}
---. 2004{\natexlab{b}}, \apj, 616, 567

\bibitem[{{Ida} \& {Lin}(2005)}]{IL05}
---. 2005, \apj, 626, 1045

\bibitem[{{Ida} \& {Lin}(2008)}]{IL08}
---. 2008, \apj, 673, 487

\bibitem[{{Ida} \& {Lin}(2010)}]{IL10}
---. 2010, \apj, 719, 810

\bibitem[{{Ida} {et~al.}(2013){Ida}, {Lin}, \& {Nagasawa}}]{IL13}
{Ida}, S., {Lin}, D.~N.~C., \& {Nagasawa}, M. 2013, \apj, 775, 42

\bibitem[{{Ikoma} \& {Genda}(2006)}]{IkomaGenda06}
{Ikoma}, M., \& {Genda}, H. 2006, \apj, 648, 696

\bibitem[{{Ikoma} {et~al.}(2000){Ikoma}, {Nakazawa}, \& {Emori}}]{Ikoma00}
{Ikoma}, M., {Nakazawa}, K., \& {Emori}, H. 2000, \apj, 537, 1013

\bibitem[{{Jennings} {et~al.}(2018){Jennings}, {Ercolano}, \&
  {Rosotti}}]{Jennings2018}
{Jennings}, J., {Ercolano}, B., \& {Rosotti}, G.~P. 2018, \mnras, 477, 4131

\bibitem[{{Kanagawa} {et~al.}(2015){Kanagawa}, {Tanaka}, {Muto}, {Tanigawa}, \&
  {Takeuchi}}]{Kanagawa15}
{Kanagawa}, K.~D., {Tanaka}, H., {Muto}, T., {Tanigawa}, T., \& {Takeuchi}, T.
  2015, \mnras, 448, 994

\bibitem[{{Kanagawa} {et~al.}(2018){Kanagawa}, {Tanaka}, \&
  {Szuszkiewicz}}]{Kanagawa18}
{Kanagawa}, K.~D., {Tanaka}, H., \& {Szuszkiewicz}, E. 2018, \apj, 861, 140

\bibitem[{{Kley} \& {Dirksen}(2006)}]{Kley06}
{Kley}, W., \& {Dirksen}, G. 2006, \aap, 447, 369

\bibitem[{{Kokubo} \& {Ida}(1998)}]{Kokubo_Ida98}
{Kokubo}, E., \& {Ida}, S. 1998, Icarus, 131, 171

\bibitem[{{Kurokawa} \& {Tanigawa}(2018)}]{Kurokawa18}
{Kurokawa}, H., \& {Tanigawa}, T. 2018, \mnras, 479, 635

\bibitem[{{Lin} {et~al.}(1996){Lin}, {Bodenheimer}, \& {Richardson}}]{Lin96}
{Lin}, D.~N.~C., {Bodenheimer}, P., \& {Richardson}, D.~C. 1996, \nat, 380, 606

\bibitem[{{Lin} \& {Papaloizou}(1986)}]{Lin1986}
{Lin}, D.~N.~C., \& {Papaloizou}, J. 1986, \apj, 309, 846

\bibitem[{{Lin} \& {Papaloizou}(1993)}]{LinPapaloizou93}
{Lin}, D.~N.~C., \& {Papaloizou}, J.~C.~B. 1993, in Protostars and Planets III,
  ed. E.~H. {Levy} \& J.~I. {Lunine}, 749--835

\bibitem[{{Lynden-Bell} \& {Pringle}(1974)}]{Lynden-Bell7}
{Lynden-Bell}, D., \& {Pringle}, J.~E. 1974, \mnras, 168, 603

\bibitem[{{Machida} {et~al.}(2010){Machida}, {Kokubo}, {Inutsuka}, \&
  {Matsumoto}}]{Machida10}
{Machida}, M.~N., {Kokubo}, E., {Inutsuka}, S.-I., \& {Matsumoto}, T. 2010,
  \mnras, 405, 1227

\bibitem[{{Mordasini} {et~al.}(2009{\natexlab{a}}){Mordasini}, {Alibert}, \&
  {Benz}}]{AMB09}
{Mordasini}, C., {Alibert}, Y., \& {Benz}, W. 2009{\natexlab{a}}, \aap, 501,
  1139

\bibitem[{{Mordasini} {et~al.}(2009{\natexlab{b}}){Mordasini}, {Alibert},
  {Benz}, \& {Naef}}]{AMB09b}
{Mordasini}, C., {Alibert}, Y., {Benz}, W., \& {Naef}, D. 2009{\natexlab{b}},
  \aap, 501, 1161

\bibitem[{{Mordasini} {et~al.}(2012){Mordasini}, {Alibert}, {Klahr}, \&
  {Henning}}]{Mordasini2012}
{Mordasini}, C., {Alibert}, Y., {Klahr}, H., \& {Henning}, T. 2012, \aap, 547,
  A111

\bibitem[{{Nagasawa} {et~al.}(2008){Nagasawa}, {Ida}, \& {Bessho}}]{Nagasawa08}
{Nagasawa}, M., {Ida}, S., \& {Bessho}, T. 2008, \apj, 678, 498

\bibitem[{{Ogihara} {et~al.}(2017){Ogihara}, {Kokubo}, {Suzuki}, {Morbidelli},
  \& {Crida}}]{Ogihara17}
{Ogihara}, M., {Kokubo}, E., {Suzuki}, T.~K., {Morbidelli}, A., \& {Crida}, A.
  2017, \aap, 608, A74

\bibitem[{{Paardekooper}(2014)}]{Paardekooper14}
{Paardekooper}, S.-J. 2014, \mnras, 444, 2031

\bibitem[{{Paardekooper} {et~al.}(2011){Paardekooper}, {Baruteau}, \&
  {Kley}}]{Paardekooper11}
{Paardekooper}, S.-J., {Baruteau}, C., \& {Kley}, W. 2011, \mnras, 410, 293

\bibitem[{{Papaloizou} \& {Larwood}(2000)}]{Papaloizou00}
{Papaloizou}, J.~C.~B., \& {Larwood}, J.~D. 2000, \mnras, 315, 823

\bibitem[{{Pinte} {et~al.}(2016){Pinte}, {Dent}, {M{\'e}nard}, {Hales}, {Hill},
  {Cortes}, \& {de Gregorio-Monsalvo}}]{Pinte16}
{Pinte}, C., {Dent}, W.~R.~F., {M{\'e}nard}, F., {et~al.} 2016, \apj, 816, 25

\bibitem[{{Rasio} \& {Ford}(1996)}]{RasioFord96}
{Rasio}, F.~A., \& {Ford}, E.~B. 1996, Science, 274, 954

\bibitem[{{Suzuki} {et~al.}(2016){Suzuki}, {Ogihara}, {Morbidelli}, {Crida}, \&
  {Guillot}}]{Suzuki16}
{Suzuki}, T.~K., {Ogihara}, M., {Morbidelli}, A., {Crida}, A., \& {Guillot}, T.
  2016, \aap, 596, A74

\bibitem[{{Szul{\'a}gyi} {et~al.}(2016){Szul{\'a}gyi}, {Masset}, {Lega},
  {Crida}, {Morbidelli}, \& {Guillot}}]{Szulagyi16}
{Szul{\'a}gyi}, J., {Masset}, F., {Lega}, E., {et~al.} 2016, \mnras, 460, 2853

\bibitem[{{Szul{\'a}gyi} \& {Mordasini}(2017)}]{Szulagyi17}
{Szul{\'a}gyi}, J., \& {Mordasini}, C. 2017, \mnras, 465, L64

\bibitem[{{Tanaka} {et~al.}(2002){Tanaka}, {Takeuchi}, \& {Ward}}]{Tanaka02}
{Tanaka}, H., {Takeuchi}, T., \& {Ward}, W.~R. 2002, \apj, 565, 1257

\bibitem[{{Tanigawa} \& {Tanaka}(2016)}]{TT16}
{Tanigawa}, T., \& {Tanaka}, H. 2016, \apj, 823, 48

\bibitem[{{Tanigawa} \& {Watanabe}(2002)}]{Tanigawa02}
{Tanigawa}, T., \& {Watanabe}, S.-i. 2002, \apj, 580, 506

\bibitem[{{Williams} \& {Cieza}(2011)}]{Williams11}
{Williams}, J.~P., \& {Cieza}, L.~A. 2011, \araa, 49, 67

\bibitem[{{Winn} {et~al.}(2010){Winn}, {Fabrycky}, {Albrecht}, \&
  {Johnson}}]{Winn10}
{Winn}, J.~N., {Fabrycky}, D., {Albrecht}, S., \& {Johnson}, J.~A. 2010, \apjl,
  718, L145

\end{thebibliography}

\end{document}